\documentclass[amsmath,amssymb,prb,twocolumn]{revtex4}

\usepackage[pdftex]{graphicx}
\usepackage{bm}
\usepackage{soul,xcolor}
\setstcolor{red}
\usepackage{pifont}
\usepackage{amssymb}
\usepackage{colordvi}
\usepackage{color}

\begin{document}
\title{
Spin-current induced mechanical torque in a chiral molecular junction
}

\author{N.~Sasao}
\address{Department of Physics Engineering, Faculty of Engineering, Mie University, Tsu, Mie, 514-8507, Japan}

\author{H.~Okada}
\address{Department of Physics Engineering, Faculty of Engineering, Mie University, Tsu, Mie, 514-8507, Japan}

\author{Y.~Utsumi}
\address{Department of Physics Engineering, Faculty of Engineering, Mie University, Tsu, Mie, 514-8507, Japan}

\author{O. Entin-Wohlman}
\affiliation{Raymond and Beverly Sackler School of Physics and Astronomy, Tel Aviv University, Tel Aviv 69978, Israel}

\author{A. Aharony}
\affiliation{Raymond and Beverly Sackler School of Physics and Astronomy, Tel Aviv University, Tel Aviv 69978, Israel}

\begin{abstract}

We analyse the appearance of a mechanical torque that acts  on a chiral molecule: a single-stranded DNA, in which the spin-orbit interaction is expected to induce a spin-selectivity effect. The mechanical torque is shown to appear  as a result of the non-conservation of the spin current in the presence of the spin-orbit interaction.
Adopting
a simple microscopic model Hamiltonian for a chiral molecule connected to source and drain leads, and
accounting for the mechanical torque acting on the chiral molecule as the back action on the electrons traversing the molecule, we derive the spin continuity-equation.
It connects the spin current expressed by a Landauer-type formula and the mechanical torque.
Thus, by injecting a spin-polarized current from the source electrode, it is possible to generate a torque, which will rotate the DNA molecule.

\end{abstract}

\date{\today}
\maketitle

\newcommand{\mat}[1]{\mbox{\boldmath$#1$}}

\section{Introduction}


Molecular junctions in which spin-orbit interactions (SOIs) are effective have  recently attracted much  attention since such interactions are considered to play an important role in electronic conduction through chiral molecules.
An  important consequence of the SOI is the spin-filter effect, in which only electrons whose  spin points along a particular direction can be transmitted. \cite{Streda2003,Pareek2004,Eto2005,Hatano2007,Aharony2011,Matityahu2017}
Recently  spin-filtering has been detected in organic chiral molecules, such as DNA molecules\cite{Goehler2011,Xie2011,Mondal2015} and peptides.\cite{Argones2017}
This effect is termed ``chiral-induced spin selectivity" (CISS).\cite{Goehler2011,Xie2011,Naaman2012} 
Experimentally, it is possible to connect one edge of a DNA molecule or a peptide molecule to a ferromagnetic electrode and the other edge to a metallic electrode through a gold nano-particle.~\cite{Xie2011,Argones2017}
Experimental results demonstrate that the magnitude  of the electric current that flows through the molecule depends on the direction of the magnetization of the ferromagnetic electrode.\cite{Xie2011,Argones2017}
Since such organic molecules do not contain magnetic atoms, it has been suggested that the only possible origin of the CISS effect would be the SOI.
In recent theoretical studies, the underlying physical mechanism of the spin polarization are studied for a double-stranded DNA~\cite{Guo2012} and for single-stranded molecules,~\cite{Guo2014,Matityahu2016} using the tight-binding model with Rashba-like SOI.
Moreover, a minimal realistic model accounting for the nature of $p$-orbitals are shown to exhibit possible CISS effect.~\cite{Varela2016}
In addition, it has suggested that the  CISS effect originates from the interaction between a helicity-induced SOI and a strong dipole electric field.~\cite{Michaeli2016,Michaeli2017}
%
%
It is also suggested that in order to verify the CISS effect in electric conduction experiments, proper multi-terminal setups are required.~\cite{Matityahu2017,Yang2019}
In order to further understand the role of the SOI in such molecules, it is of interest to discuss other possible consequences.

Quite generally,  the SOI is described by the Hamiltonian
\begin{align}
{\cal H}^{}_{\rm SOI} \propto ({\bm E} \times {\bm p} ) \cdot {\bm \sigma}
\ ,
\end{align}
which implies an interplay among the electron spin $\bm{s}=(\hbar/2){\bm \sigma}$ [$
{\bm \sigma}
=
(
\sigma_x,
\sigma_y,
\sigma_z
)
$
is the vector of the Pauli matrices], the electron momentum ${\bm p}$ and the electric field ${\bm E}$.
This electric field is generated by the nuclei forming the molecule, and therefore  its dynamics is connected with the dynamics of the molecule.
In addition, since the atomic orbitals are attached to the nuclei of the molecule,  orbital electronic angular momentum can be transformed into mechanical angular momentum of the molecule.
This observation implies the SOI-assisted conversion of the spin angular momentum into a mechanical angular momentum.
In ferromagnetic materials, such a conversion, i.e., the gyromagnetic effect, has been known already a century  ago.\cite{Chikazumi,Richardson1908,Einstein1915,Barnett1915}
The conversion between the spin angular momentum and the mechanical torque has received recently renewed attention in the spintronics community.\cite{Fulde1998,Mohanty2004,Malshukov2005,Kovalev2007,Matsuo2013,Matsuo2014,Matsuo2015,Matsuo2017}
In the presence of the SOI, the direction of an injected spin varies continuously during the transmission process.
Then a certain  amount of  angular momentum is transferred to the atoms in the form of back action.
Therefore, one can expect that in molecular junction setups, \cite{Xie2011,Argones2017} a finite amount of  spin angular momentum will be converted into a mechanical torque, which in turn will operate on the chiral molecule. This operation could lead to a rotational motion of the chiral molecule.

In the present paper we discuss the mechanical torque induced by a spin-polarized electric current.
Because of the SOI, the total spin of electrons is not conserved, as opposed to the total angular momentum. Consequently the change in the electron spin is transformed into a mechanical torque acting on the molecule. Assuming for simplicity that the system is at zero temperature, we
derive a continuity equation for the spin,  based on a microscopic model Hamiltonian for chiral molecules introduced in previous publications. \cite{Guo2012,Guo2014,Matityahu2016} At steady state, this equation relates the spin current to the mechanical torque acting on the chiral molecule. The dependence of this mechanical torque on the parameters of the molecule is the central issue of our discussion.

Our paper is structured as follows. We begin in Sec. \ref{sec:hamiltonian} by presenting the model Hamiltonian for the single-stranded DNA molecule. In particular we elaborate on the effect of the spin-orbit interaction on the Hamiltonian of the molecule and the tunnel couplings to the left and right electrodes. We then continue in Sec.
\ref{sec:spin_continuity} to construct the operator form of the continuity equation for the spin. This equation, which is
based on the Euler-Lagrange
equation for the rotation angle of the molecule,
 involves the spin-polarized current operator and the operator of the mechanical torque. The quantum average of the operator continuity equation is considered  in Sec.
\ref{sec:Landauer}. There, we first construct   the scattering states of spin-polarized electrons impinging on the molecule (details are given in Appendix \ref{sec:detailed_Landauer}) and then exploit those to perform the quantum average. As a result, we obtain a Landauer-type expression for the  spin-polarized currents. These currents determine the mechanical torque acting on the molecule at steady state.
Our results are presented in Sec.
\ref{sec:results}. We display there the mechanical torque as a function of the energy of the impinging electrons, and as a function of the
structural parameters of the molecule, in a range of values of the SOI. The most conspicuous feature of the data presented in Sec. \ref{sec:results} are the periodic oscillations of the mechanical torque (see Figs.
\ref{fig:t_vs_e}, \ref{fig:t_vs_n},  and \ref{fig:t_vs_n2}). From the various analyses of our numerical data it emerges that the periodicity and amplitude of the oscillations originate from an interplay between the spin-orbit interaction active in the molecule, and the chiral structure of the latter.
This observation is substantiated by calculating the band structure of our model for the molecule, as detailed in Appendix
\ref{sec:energy_band}.
Our findings are summarized in Sec.
\ref{sec:summary}.

\section{Mechanical torque induced by spin currents}
\label{SecMec}

\subsection{Single-stranded DNA model Hamiltonian}
\label{sec:hamiltonian}

\begin{figure}[ht]
\includegraphics[width=.75 \columnwidth]{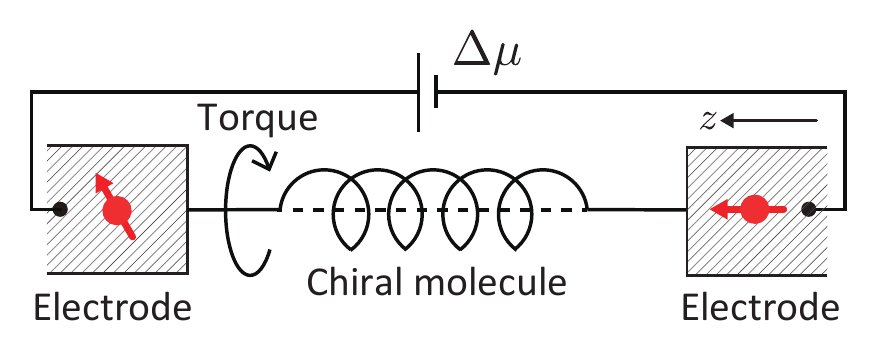}
\caption{
A single-stranded DNA molecule, whose two edges are connected to left and right electrodes.
A spin-polarized electron injected from the left lead changes its spin direction during the passage through the molecule due to   the spin-orbit interaction active there. The back action of this process results in a
mechanical torque acting on the molecule.
 The rotation axis of the chiral molecule is along the $\hat{\bm z}$ axis; $\Delta \mu$ is the chemical-potential difference between the two electrodes.}
\label{fig:setup}
\end{figure}
Figure \ref{fig:setup} displays schematically the molecular junction.
The two edges of the chiral molecule, a single-stranded DNA molecule, are connected to  left and right leads.
The  Hamiltonian of the system  consists of the DNA  Hamiltonian,  $\mathcal{H}_{\rm{mol} }$, the left (right) lead Hamiltonian,  $\mathcal{H}_{L(R)}$,  and the tunneling Hamiltonian $V$,
\begin{align}
\mathcal{H} = \mathcal{H}^{}_{\rm{mol} } + \mathcal{H}^{}_{L} + \mathcal{H}^{}_{R} + V
\, .
\label{eq:Hamiltonian}
\end{align}

\begin{figure}[ht]
\includegraphics[width=.75 \columnwidth]{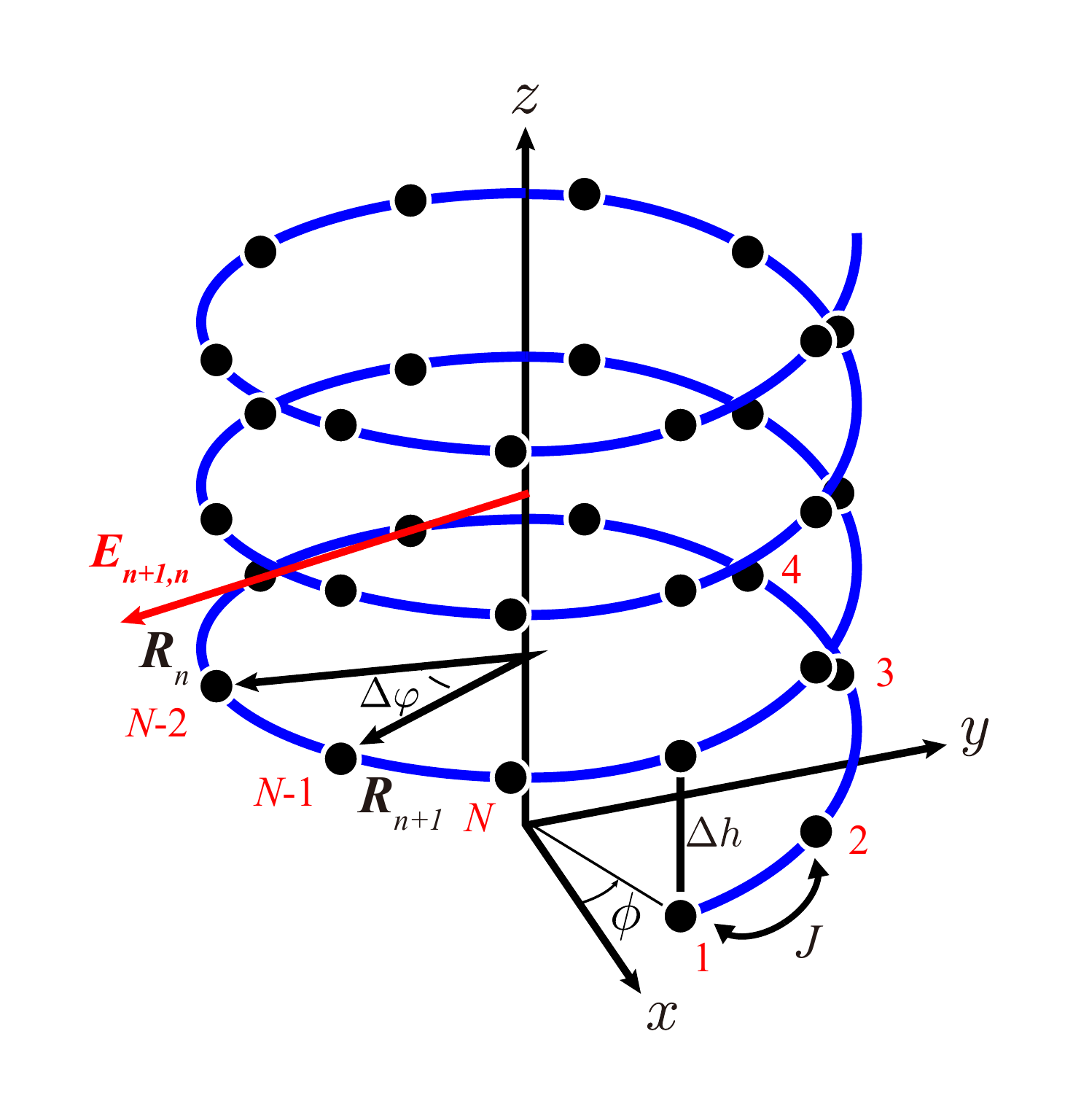}
\caption{
The effective model of the single-stranded DNA:
 a helix made of a one-dimensional tight-binding chain with radius $R$ and pitch $\Delta h$.
The total number of  sites in this figure is $N_{\rm mol}=32$ and the number of sites in a unit cell is $N=10$. The angles $\Delta\varphi$ and $\phi$ are explained in Eqs. (\ref{eq:R}) and the following text.
}
\label{fig:dna}
\end{figure}
We model the single-stranded DNA molecule by a helix of a one-dimensional tight-binding chain propagating chirally around a cylinder of radius $R$ whose axis is along the $\hat{\bm z}$ direction, with a pitch $\Delta h$ in the counterclockwise direction (Fig.~\ref{fig:dna}).
We assume that each unit cell contains $N$ sites. Counting from the left electrode (lowest in Fig. ~\ref{fig:dna}), the
$n$-th site of the chain is located at
\begin{align}
{\bm R}^{}_{n} =&
R [ \cos(\varphi_n+\phi) \hat{{\bm x}} + \sin(\varphi_n+\phi) \hat{{\bm y}}]
 +
\Delta h \frac{\varphi_n}{2 \pi} \hat{{\bm z}}
\ ,\nonumber\\
\varphi_n =& n \, \Delta \varphi \, ,
\;\;\;\;
\Delta \varphi = \frac{2 \pi}{N} \, .
\label{eq:R}
\end{align}
Here  $\hat{{\bm x}}=(1,0,0)$, $\hat{{\bm y}}=(0,1,0)$ and $\hat{{\bm z}}=(0,0,1)$ are unit  vectors along the cartesian axes.
The  angle $\phi$   by which the entire molecule rotates,  is the dynamical variable corresponding to  the rotary motion of the molecule.
The SOI active in the chain  is modeled by an Aharonov-Casher phase~\cite{Aharonov1984,Aronov1993} emerging due to an electric field generated by an imaginary line of charge situated along the $\hat{\bm z}$ axis.
Because of the SOI, the spin of the electron rotates as it  tunnels from the $n$th site to the $(n+1)$th site. This rotation is  described by a $2 \times 2$ unitary matrix,
\begin{align}
V^{}_n=e^{ i\bm{K}_{n,n+1}  \cdot \bm{\sigma} }= \cos(\alpha) +i \sin(\alpha) \hat{\bm {K}}^{}_{n,n+1}  \cdot \bm{\sigma}
\, ,
\label{eqn:V_n}
\end{align}
where $\hat{\bm {K}}^{}_{n,n+1}$ is a unit vector along the direction of the rotation vector $
\bm{K}_{n,n+1}$. 
The vector ${\bm K}_{n+1,n}$ is given by
\begin{align}
{\bm K}^{}_{n+1,n} =
\lambda \,
({\bm R}^{}_{n+1} - {\bm R}^{}_{n})
\times
{\bm E}^{}_{n+1,n}
\, ,
\label{eq:K_n}
\end{align}
where  $\lambda$  parameterizes the strength of the SOI,  and
${\bm E}_{n+1,n}$ is the electric field which is perpendicular to the $\hat{\bm z}$ axis at the midpoint between the $n$th site and the $(n+1)$th site, $({\bm R}_{n+1} + {\bm R}_n)/2$,
\begin{align}
{\bm E}^{}_{n+1,n}
=
E^{}_0
\left[
\cos ( \varphi^{}_{n+1/2} +\phi ) \hat{{\bm x}}
+
\sin ( \varphi^{}_{n+1/2} +\phi ) \hat{{\bm y}}
\right]
\  .
\label{E_n}
\end{align}
The magnitude  of the SOI in our model,
\begin{align}
\alpha
=\left| {\bm K}^{}_{n+1,n} \right|
=
\lambda E^{}_0
\sqrt{
4R^2
\sin^2 \left(\frac{\pi}{N}\right)
+
\left(\frac{\Delta h}{N} \right)^2
}
\, ,
\end{align}
is independent of the position $n$.
Assuming that  the DNA molecule contains $N_{\rm mol}$ sites, the model Hamiltonian  we use
 is~\cite{Guo2012,Guo2014,Matityahu2016}
\begin{align}
 \mathcal{H}^{}_{\rm mol} = \sum_{n=1}^{N_{\rm mol}} \epsilon^{}_n c_{n}^\dagger c^{}_{n} -J \sum_{n=1}^{{N_{\rm mol}}-1} c_{n+1}^\dagger V^{}_n c^{}_{n}  + {\rm H.c.} \ .
\label{eq:H_mol}
\end{align}
Here, $\epsilon_n$ is the on-site energy, taken below as zero for simplicity, and  $J$ is the hopping amplitude (in energy units) between  nearest-neighbor sites.
The Hamiltonian of  lead $r$ ($r=L,R$) is
\begin{align}
\mathcal{H}_{r} = -J_0 \sum_{n=1}^{N_r-1} c_{r,n+1}^\dagger c^{}_{r,n} +  {\rm H.c.}
\ ,
\label{eq:H_R}
\end{align}
where $N_r$ is the number of sites in the $r$ lead, taken below to be infinity.
All annihilation and creation operators [see, Eqs. (\ref{eq:H_mol}), (\ref{eq:H_R}), and (\ref{eqn:v})] are spinors; e.g.,
\begin{align}
c_{n}
=
\left(
\begin{array}{c}
c_{n \uparrow} \\
c_{n \downarrow}
\end{array}
\right)
\,\, ,
c_{r,n}
=
\left(
\begin{array}{c}
c_{r,n \uparrow} \\
c_{r,n \downarrow}
\end{array}
\right)
\, .
\end{align}
The tunneling Hamiltonian between the leads and the molecule is
\begin{align}
V = v c_{1}^\dagger c^{}_{R,1} + v c_{N_{\rm mol}}^\dagger c^{}_{L,1}+  {\rm H.c.}
\ .
\label{eqn:v}
\end{align}
It is assumed  that an electron at the 1st site of the right (left) lead can  tunnel only to the 1st ($N_{\rm mol}$-th) site of the molecule.

\subsection{Spin continuity-equation}
\label{sec:spin_continuity}

The  continuity equation of the spin can be derived within a semi-classical approximation.
The time scale of the electron dynamics is much shorter than that of the ions comprising the chiral molecule.
Therefore, we first fix the rotation angle $\phi$ of the molecule and  consider the eigen wave-function $|\psi(\phi) \rangle$.   The Lagrangian of the molecule is then \cite{LandauLifshitz}
\begin{align}
{\mathcal L}_{{\rm mol} } = \frac{I}{2} \dot{\phi}^2 - \langle \psi(\phi) |\mathcal{H} |\psi(\phi) \rangle \ .
\end{align}
Here
$I = N_{\rm mol} \, m_{\rm I} R^2$ is the
 moment of inertia of the DNA ions,
and $m_{\rm I}$ is the mass of a single  ion.
The dynamics of the rotation angle $\phi$ is governed by the Euler-Lagrange equation:
\begin{equation}
\frac{d}{dt} \left( \frac{\partial {\mathcal L}_{{\rm mol} }}{\partial \dot{\phi}} \right)
+ \frac{\partial {\mathcal L}_{{\rm mol} }}{\partial \phi} =0
\, ,
\end{equation}
then, 
\begin{align}
I \ddot{\phi} = \langle \psi(\phi) |
\frac{\partial \mathcal{H}_{\rm mol}(\phi)}{\partial \phi}
|\psi(\phi) \rangle
\equiv \tau
\ ,
\label{eqn:eleq}
\end{align}
with the mechanical torque $\tau$  given by the quantum average of the operator of the mechanical torque [see
Eqs. (\ref{eqn:V_n}), (\ref{eq:K_n}),  and (\ref{E_n})],\cite{AAcom}
\begin{align}
&\frac{\partial \mathcal{H}_{\rm mol}(\phi)}{\partial \phi}
= -i J \sin (\alpha)\nonumber\\
&\times\Big ( \sum_{n=1}^{N_{\rm mol}-1} c_{n+1}^\dagger [\hat{\bm K}^{}_{n+1,n} \times {\bm \sigma}]^{}_z  c^{}_{n}
- {\rm H.c.}\Big )
\ .
\label{eq:H_mol_torque}
\end{align}
We next relate the operator of the mechanical torque,
$\partial \mathcal{H}_{\rm mol}(\phi)/(\partial \phi)$, to the spin-current operator. This is accomplished as follows.
The operator of the  total electrons' spin on the molecule,
${\bm S}_{\rm mol}$,
 along an arbitrary direction of a unit vector $\hat{{\bm \ell}}$,
reads
\begin{align}
{\bm S}_{\rm mol} \cdot \hat{{\bm \ell}} = \frac{\hbar}{2} \sum_{n=1}^{N_{\rm mol}} c_{n}^\dagger ( \hat{{\bm \ell}} \cdot {\bm \sigma} ) c^{}_{n} \ .
\label{eq:S_mol}
\end{align}
Exploiting  the
Heisenberg picture  in which
${\bm S}_{\rm mol}(t)=\exp[i \mathcal{H} (t-t_0)/\hbar ]{\bm S}_{\rm mol} \exp[-i \mathcal{H}(t-t_0)/\hbar]$,
we derive the equation of motion
\begin{align}
\frac{d}{dt}
[{\bm  S}_{\rm mol}(t) \cdot \hat{{\bm \ell}}]
&=
I^{}_{L,\hat{{\bm \ell}}}(t) + I^{}_{R,\hat{{\bm \ell}}}(t)+{\bm J}\cdot\hat{{\bm \ell}} \ .
\label{eq:dS_mol_dt}
\end{align}
The first and  second terms on the right hand-side of Eq.~(\ref{eq:dS_mol_dt}) are the operators for  the spin currents in the left and right interfaces,
\begin{align}
I_{L(R),\hat{{\bm \ell}}} = \frac{iv}{2} \left( c_{L(R),1}^\dagger  ( \hat{{\bm \ell}} \cdot {\bm \sigma} ) c^{}_{1(N_{\rm mol})}-{\rm H.c.}
\right)
\ .
\label{eq:sco}
\end{align}
The third term there
corresponds to the source of spin current that flows under the effect of the SOI,
\begin{align}
&{\bm J}\cdot \hat{{\bm \ell}}=
-i J \sin (\alpha) \nonumber\\
&\times\Big (\sum^{N_{\rm mol}-1}_{n=1} c^{\dagger}_{n+1} [\hat{{\bm \ell}} \cdot (\hat{\bm K}^{}_{n+1,n} \times {\bm \sigma})] c^{}_n
-
{\rm H.c.}\Big )
\  .
\label{eq:source}
\end{align}
Indeed, comparing Eq. (\ref{eq:H_mol_torque}) for the operator of the mechanical torque with
Eq.
(\ref
{eq:source}), one concludes that the former is just the $z-$component of ${\bm J}$.
Hence,
\begin{align}
\frac{d {\bm S}^{}_{{\rm mol},z}(t)}{dt}
&=
\frac{\partial \mathcal{H}_{\rm mol}(\phi)}{\partial \phi}
+
I^{}_{L,\hat{\bm z}}(t) + I^{}_{R,\hat{\bm z}}(t)
\, ,
\label{eq:spin_conti}
\end{align}
 is the spin continuity-equation:  the derivative $\partial \mathcal{H}_{\rm mol}(\phi)/\partial \phi$ acts as the source of the spin angular-momentum of the electron.
The other components of the spin-current source, i.e.,  $J_x$ and $J_y$,  are not conserved, because
the two edges'  couplings  to the left and right leads prevent the molecule from rotating along the $\hat{\bm x}$ and $\hat{\bm y}$ axes. (Obviously the interfaces' currents $I_{L,\hat{\bm z}}$ and $I_{R,\hat{\bm z}}$ flow along the $\hat{\bm z}-$direction.)

At steady state  the time derivative $d {\bm S}_{{\rm mol},z}(t)/dt $ vanishes, and consequently [see Eq. (\ref{eqn:eleq})]
\begin{align}
I \ddot{\phi} = - \langle \psi (\phi)| (I_{L,\hat{\bm z}}  + I_{R,\hat{\bm z}}) |\psi (\phi) \rangle = \tau \ .
\label{eq:spin_conti_eq}
\end{align}
This is the equation of motion of a rotating rigid body: it relates the spin current with the mechanical torque $\tau$.
A similar equation of motion, derived
phenomenologically from the conservation of the total angular momentum,  has been found in  Ref.~\onlinecite{Mohanty2004} [see in particular Eq. (5) there]. In contrast, the continuity equation (\ref{eq:spin_conti_eq})  is
derived form a specific microscopic Hamiltonian within the semi-classical approximation.
The considerations given in this section  are in parallel with those related to the  current-induced spin transfer torque: \cite{Slonczewski1996} a spin-polarized current that generates a torque which induces in turn  background magnetic moments. In our case, however,  the spin angular-momentum is converted into a mechanical torque      due to the relativistic correction brought about by the SOI in the non-relativistic Schr\"odinger equation.


\subsection{Landauer-type formula}
\label{sec:Landauer}

To complete the calculation of the continuity equation at steady state, Eq. (\ref{eq:spin_conti_eq}), we need to find the quantum average of the spin currents $I_{L,\hat{\bm z}}$ and $I_{R,\hat{\bm z}}$. This is accomplished within  the scattering formalism. We
replace $|\psi(\phi) \rangle$ by the corresponding  scattering state,    obtained upon  adiabatically switching-on the tunneling Hamiltonian $V$. In this way, $\langle \psi(\phi)| I^{}_{L(R),\hat{\bm z}}|\psi(\phi)\rangle $ attains the form of a Landauer-type formula.

The wave function of an electron, of wave number $2\pi \ell/N_r$ ($\ell=1,\cdots,N_r-1$),  in dimensionless units and spin $\sigma$,
 impinging on the molecule from the right lead is
 $| \ell \sigma \rangle_R = c_{R, \ell \sigma}^\dagger |0 \rangle$,
where $|0 \rangle$ is the vacuum state and
$c_{R,\ell \sigma}^\dagger $ is the   creation operator
\begin{align}
c_{R,\ell \sigma}^\dagger = \sqrt{ \frac{2}{N^{}_R} } \sum_{n=1}^{N^{}_R}
\sin \left( 2 \pi \frac{n \ell}{N^{}_R} \right) c_{R,n \sigma}^\dagger \ .
\label{cks}
\end{align}
In the  $N_R \to \infty$ limit,
$| \ell \sigma \rangle_R$
 is an eigenstate  of the decoupled right lead,
$\mathcal{H}_R | \ell \sigma \rangle_R = E_{R \ell \sigma} | \ell \sigma \rangle_R$,
with the eigen energy
\begin{align}
E^{}_{R,\ell \sigma} = -2 J^{}_{0} \cos  \left( 2 \pi \frac{\ell}{N^{}_R} \right)
\ .
\end{align}
The scattering state excited by an electron in the eigenstate $| \ell \sigma \rangle_R$ is the solution of  the Lippmann-Schwinger equation, \cite{JJSakurai}
\begin{align}
| \ell \sigma^{(+)}  (\phi) \rangle^{}_R
=
[1 + G (E_{R,\ell\sigma} + i0) V ]
| \ell \sigma \rangle^{}_R
\ ,
\label{eq:ss}
\end{align}
where the Green's function operator $G(E)$ is
\begin{align}
G(E)
=
[E-\mathcal{H}]^{-1}
\ .
\end{align}
The dependence on the rotation angle $\phi$  of the scattering wave function emerges from that of  the Hamiltonian in the Green's function operator.
When the quantum average of the spin currents is calculated with the scattering  states of  an electron with spin $\sigma$ and energy $E$  injected  from the right lead (see Appendix~\ref{sec:detailed_Landauer} for details), the mechanical torque $\tau$, Eq.~(\ref{eq:spin_conti_eq}), becomes
\begin{align}
\tau = & \sum_{\ell} {}^{}_R\langle \ell \sigma^{(+)}  (\phi) | (I_{L,\hat{\bm z}}  + I_{R,\hat{\bm z}}) | \ell \sigma^{(+)} (\phi) \rangle^{}_R 
\nonumber \\ 
& \times \delta(E^{}_{R,\ell \sigma}-E) \Delta \mu
\nonumber\\ 
= &
\Delta \mu
\big [
-g^{}_{R \downarrow, R \uparrow} -g^{}_{L \downarrow, R \uparrow}
+g^{}_{R \uparrow, R \downarrow} +g^{}_{L \uparrow, R \downarrow}
\big]\ ,
\label{eq:st}
\end{align}
where $\Delta \mu/e$ is the source-drain bias voltage (see Fig.~\ref{fig:setup}), and where the first (last) two terms on the RHS come from $\sigma=\uparrow$  ($\sigma=\downarrow$). Equation (\ref{eq:st}) is valid in the linear-response regime and  at zero temperature.
The spin-mixing spin-conductance,
$g_{r \sigma, r' \overline{\sigma}}$ [
$\overline{\sigma}=\uparrow (\downarrow)$ for $\sigma=\downarrow (\uparrow)$],
is
\begin{align}
g^{}_{r \sigma, r' \overline{\sigma}}(E) = \pi
\rho{}^{}_{r \sigma}(E)
| ^{}_r\langle 1 \sigma| T(E) | 1 \overline{\sigma} \rangle_{r'} |^2
\rho_{r' \bar{\sigma}}(E)
\, ,
\label{eq:sm-sc}
\end{align}
where
$| 1 \sigma \rangle_r = c_{r,1 \sigma}^\dagger |0 \rangle$, and $r=L,R$.
The $T$-matrix operator in Eq. (\ref{eq:sm-sc})
is given by~\cite{JJSakurai}
\begin{align}
T(E) = V +  V G(E) V
\, ,
\label{eq:t_matrix}
\end{align}
and the local density of states is
\begin{align}
\rho^{}_{r \sigma}(E)
=
\sum_\ell \left| {}^{}_r\langle 1 \sigma| \ell \sigma \rangle^{}_r \right|^2
\delta(E-E^{}_{r,\ell \sigma})
\ .
\label{eq:LDOS}
\end{align}
From Eq. (\ref{cks}), $\left| {}^{}_r\langle 1 \sigma| \ell \sigma \rangle^{}_r \right|^2=(2/N^{}_R)\sin^2(\pi \ell/N^{}_R)$.
Equation~(\ref{eq:st}) for the mechanical torque  is a Landauer-type formula.
It indicates that the transmission and reflection processes which are accompanied by spin  flipping of the transferred electrons  determine the spin current, and in turn, the induced mechanical torque.  This is similar to the case of the current-induced spin transfer torque. ~\cite{Slonczewski1996,Utsumi2015}

In order to facilitate our  numerical calculations [see Sec. \ref{sec:results}], we  rewrite the spin-mixing spin-conductance Eq.~(\ref{eq:sm-sc}) as
\begin{subequations}
\begin{align}
g^{}_{L \sigma, L \overline{\sigma}}(E) =& \frac{1}{\pi} \, {\rm Im} \Sigma_{L \sigma}(E+i0) \, {\rm Im} \Sigma_{L \overline{\sigma}}(E+i0)
\nonumber
\\
& \times
\left| \langle N_{\rm mol} \sigma| G_{\rm mol}(E+i0) | N_{\rm mol} \overline{\sigma} \rangle \right|^2
\, ,
\\
g^{}_{R \sigma, R \overline{\sigma}}(E) =& \frac{1}{\pi} \, {\rm Im} \Sigma_{R \sigma}(E+i0) \, {\rm Im} \Sigma_{R \overline{\sigma}}(E+i0)
\nonumber
\\
& \times
\left| \langle 1 \sigma| G_{\rm mol}(E+i0) | 1 \overline{\sigma} \rangle \right|^2
\, ,
\\
g^{}_{L \sigma, R \overline{\sigma}}(E) =& \frac{1}{\pi} \, {\rm Im} \Sigma_{L \sigma}(E+i0) \, {\rm Im} \Sigma_{R \overline{\sigma}}(E+i0)
\nonumber
\\
& \times
\left| \langle N_{\rm mol} \sigma| G_{\rm mol}(E+i0) | 1 \overline{\sigma} \rangle \right|^2
\, ,
\\
g^{}_{R \sigma, L \overline{\sigma}}(E) =& \frac{1}{\pi} \, {\rm Im} \Sigma_{R \sigma}(E+i0) \, {\rm Im} \Sigma_{L \overline{\sigma}}(E+i0)
\nonumber
\\
& \times
\left| \langle 1 \sigma| G_{\rm mol}(E+i0) | N_{\rm mol} \overline{\sigma} \rangle \right|^2
\, .
\end{align}
\end{subequations}
The Green's function operator for the  molecule and the corresponding self-energy are
\begin{align}
G^{-1}_{\rm mol}(E)
=&
E - \mathcal{H}^{}_{\rm mol} -\sum_{\sigma} \Sigma^{}_{L \sigma} (E) |N_{\rm mol} \sigma \rangle \langle N_{\rm mol} \sigma|
\nonumber\\
&+
\Sigma^{}_{R \sigma} (E)
|1 \sigma \rangle \langle 1 \sigma|
\ ,
\label{eq:res_red}
\end{align}
and
\begin{align}
\Sigma^{}_{r \sigma}(E)=&v^2 G^{}_{r \sigma}(E)
\ ,
\label{eq:selfe}
\end{align}
respectively [$v$ is the tunneling amplitude between the leads and the molecule, see  Eq.~(\ref{eqn:v})].
In the limit  $N_r \to \infty$, the boundary Green's function of  lead $r$ can be  calculated analytically,
\begin{align}
G^{}_{r \sigma}(E+i0)
=&
{}^{}_r\langle 1 \sigma|[E+i0-\mathcal{H}^{}_r]^{-1}| 1 \sigma \rangle^{}_r
\nonumber\\
=&
\frac{1}{J_0}
\left \{
\begin{array}{cc}
\varepsilon -   \sqrt{\varepsilon^2 -1} & (\varepsilon >1) \\
\varepsilon - i \sqrt{1- \varepsilon^2} & (|\varepsilon| \leq 1) \\
\varepsilon +   \sqrt{\varepsilon^2 -1} & (\varepsilon <-1)
\end{array}
\right.
\, ,
\end{align}
where $\varepsilon = E/(2 J_0)$.
The local density of states is $\rho_{r \sigma } (E) = -{\rm Im} G_{r \sigma}(E+i0)/\pi$.
The  inverse matrix in Eq.~(\ref{eq:res_red}) is computed numerically.

Several comments are in order.
There would be the opposite effect, namely an external torque applied to the molecule would generate a spin current. 
The rotary motion would pump spin current~\cite{Tserkovnyak2002} and dissipate the mechanical angular momentum, which could introduce a damping term in Eq.~(\ref{eq:spin_conti_eq}). 
Equation~(\ref{eq:spin_conti_eq}) also does not account for the effect of nonequilibrium shot noise.
Due to electric-current fluctuations, the spin current will also fluctuates, which would result in  fluctuations in the mechanical torque as well.
In order to take effects of nonequilibrium shot noise and damping into account, one needs to adopt the theory of full-counting statistics.\cite{Utsumi2015}
Another intriguing  outcome of our model is that the spin-mixing spin-conductance Eq.~(\ref{eq:sm-sc}) is independent of the rotation angle $\phi$, and consequently the torque is also independent of that angle.
This can be proven as follows.
By exploiting the property of the unitary matrix Eq.~(\ref{eq:V_n_}), i.e.,
$V_n(\phi) = \exp[-i \phi \sigma_z /2]V_n(\phi=0) \exp[i \phi \sigma_z /2]$,
the Hamiltonian of the molecule can be written as ${\mathcal H}_{\rm mol}(\phi) = \exp[-i \phi \sigma_z /2] {\mathcal H}_{\rm mol}(\phi=0) \exp[i \phi \sigma_z /2]$.
It follows that the $T$-matrix operator, Eq.  (\ref{eq:t_matrix}),  attains the same form, $T(E;\phi) = \exp[-i \phi \sigma_z /2] T(E;\phi=0) \exp[i \phi \sigma_z /2]$,
and thus its matrix element acquires a  phase factor due to the rotation:
$^{}_r\langle 1 \sigma| T(E;\phi) | 1 \overline{\sigma} \rangle_{r'}= {}^{}_r\langle 1 \sigma| T(E;\phi=0) | 1 \overline{\sigma} \rangle_{r'} \exp[-i \sigma \phi]$.
Inserting this form into Eq.~(\ref{eq:sm-sc}) verifies that the  spin-conductance and the torque are  independent of the rotation angle, that is
$g_{r \sigma, r' \bar{\sigma}}(E;\phi)=g_{r \sigma, r' \bar{\sigma}}(E;\phi=0)$, and $\tau(\phi)=\tau(\phi=0)$.

\section{results and discussions}
\label{sec:results}

\begin{figure}[ht]
\includegraphics[width=.85 \columnwidth]{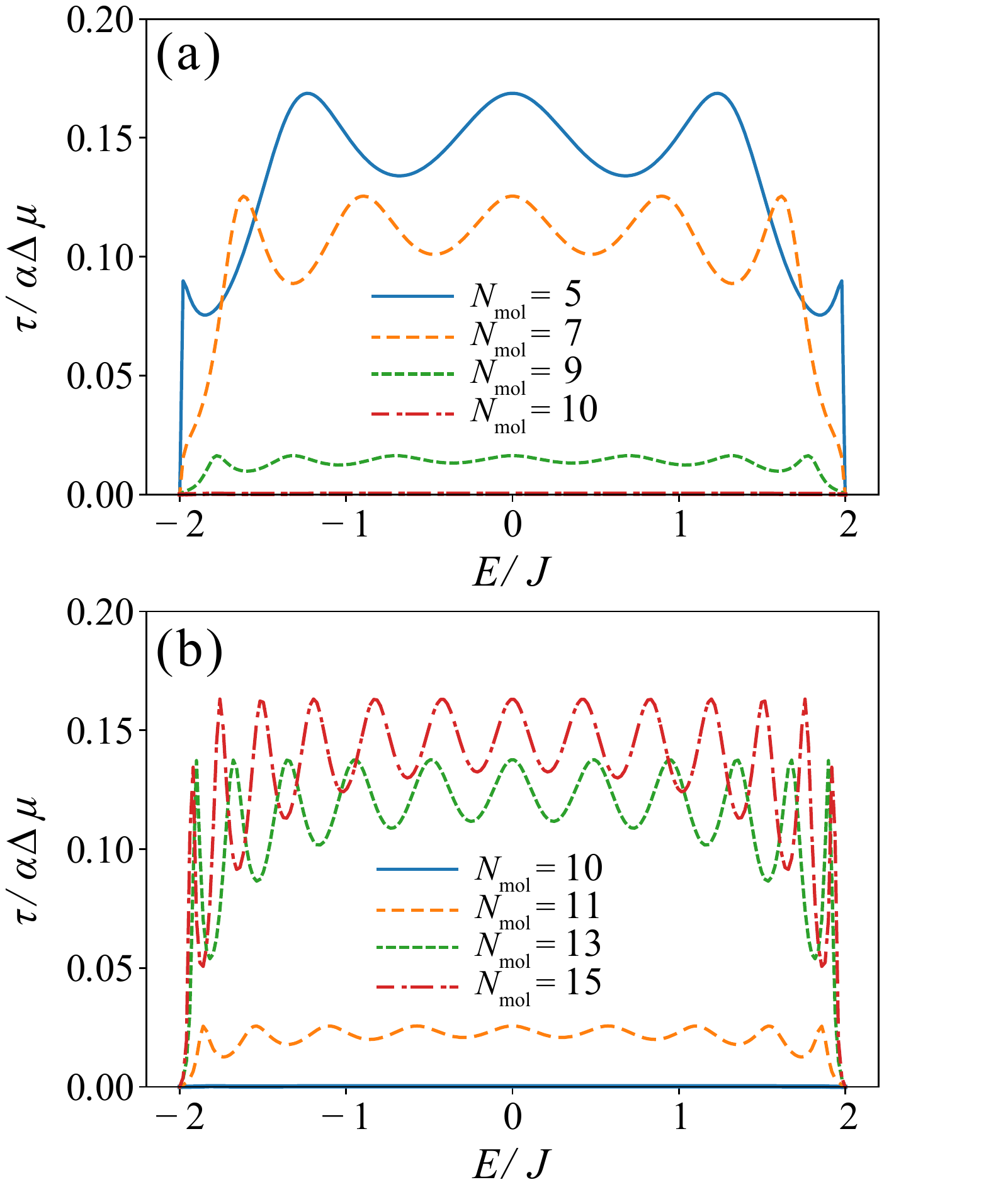}
\caption{
The mechanical torque in units of the spin conductance (i.e., normalized by the strength of the SOI), as a function of the energy of the injected electron, for several values of the number of sites  on the molecule, $N_{\rm mol}$.
The spin of the injected electron points along the $\hat{\bm z}$ direction.
There are $N=10$ sites in each turn of the DNA helix.
Panel (a) is for $N_{\rm mol}=5,7,9,10$, i.e., the molecule is shorter than its unit cell  and panel (b) is for $N_{\rm mol}=10,11,13,15$, longer than the unit cell. The parameters are:
$J_0=J$, $v=\sqrt{1.6} J$, and $\Delta h/R = 18.1$.
}
\label{fig:t_vs_e}
\end{figure}

Figures \ref{fig:t_vs_e} show the mechanical torque induced by electrons injected with their spins polarized along the $\hat{\bm z}-$direction,  as a function of their energy $E$.
In panel (a) the length of the molecule (i.e., the number of sites, $N_{\rm mol}$)  is shorter than  the unit cell ($N_{\rm mol} \leq N=10$).
Panel (b) displays the dependence of the mechanical torque on the energy for molecule's lengths   longer than the length of the unit cell,  $N_{\rm mol} \geq N$.
The curves in both panels reveal that the mechanical torque oscillates as a function of the energy, and reaches a maximal value of about
$\tau \sim 0.15 \alpha \Delta \mu$.
The oscillations' periodicity  decreases as the length $N_{\rm mol}$  increases.
Moreover, the overall amplitude of the torque  oscillates as well with the  increasing length.
For $N_{\rm mol}=10$, the mechanical torque is almost vanished, see Fig.~\ref{fig:t_vs_e} (a) and (b).

\begin{figure}[ht]
\includegraphics[width=1. \columnwidth]{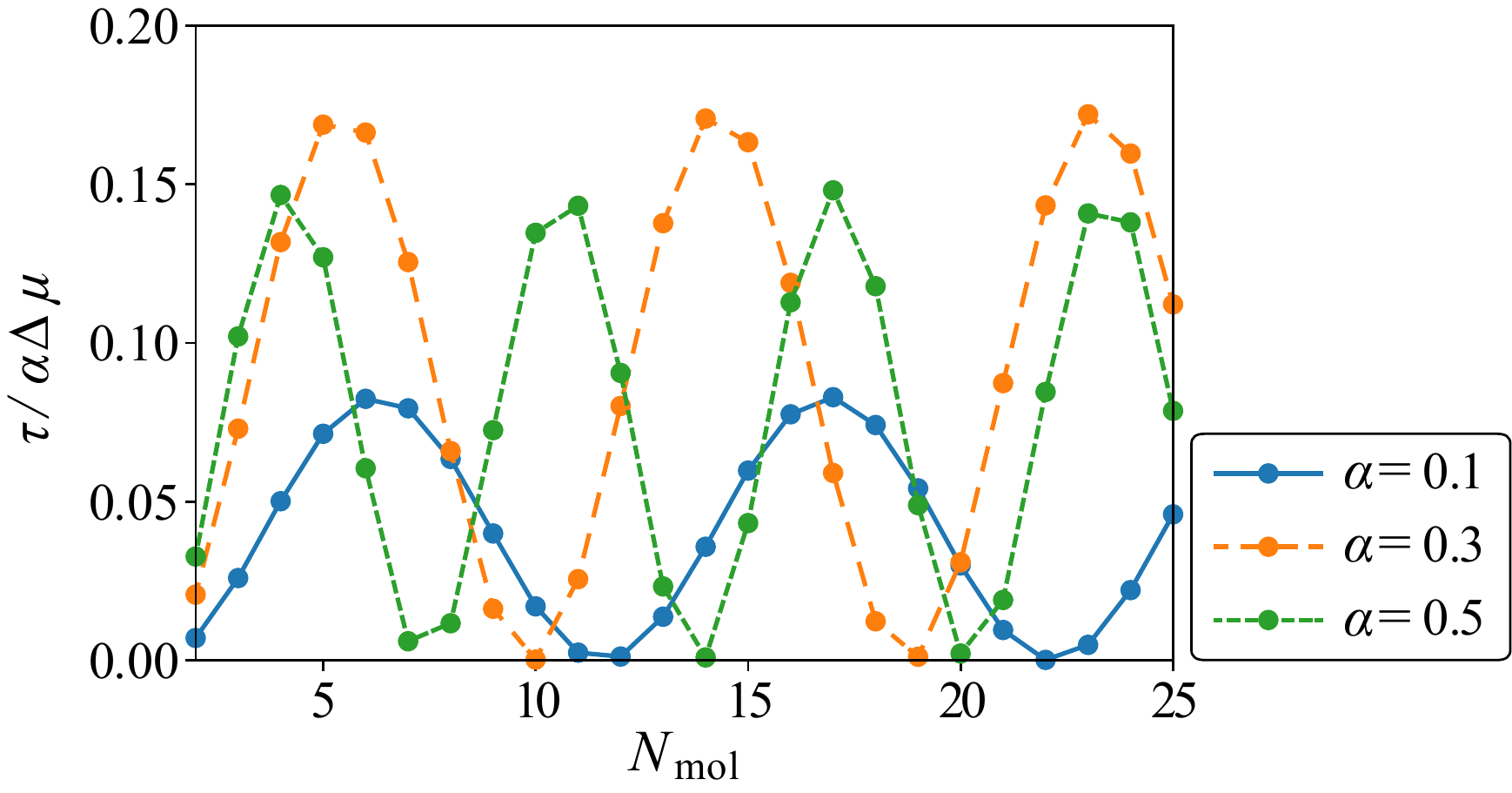}
\caption{
The normalized torque as a function of the length of the molecule (i.e., the number of sites, $N_{\rm mol}$), for several values of the spin-orbit interaction strength, $\alpha=0.1,0.3,0.5$.
With the  increase of the SOI  strength, the amplitude becomes larger while the oscillation period becomes shorter.
The energy of the injected electrons is chosen to be  $E=0$.
The other parameters are  as in Fig. \ref{fig:t_vs_e}.
}
\label{fig:t_vs_n}
\end{figure}

Figure~\ref{fig:t_vs_n} exhibits the mechanical torque as a function of the length of the molecule $N_{\rm mol}$ for several values of the strength of the SOI,
with the torque normalized by the strength. The torque oscillates as a function of the length of the molecule.
One notes that with the increase of the SOI strength, the normalized amplitude becomes larger while the oscillation period becomes shorter.
It implies that the oscillation period is determined by the strength of the SOI.
On the other hand, for weaker SOI couplings, the oscillation period approaches  the length of the unit cell,  $N=10$ in the case of Fig. \ref{fig:t_vs_n}.

\begin{figure}[ht]
\includegraphics[width=0.95 \columnwidth]{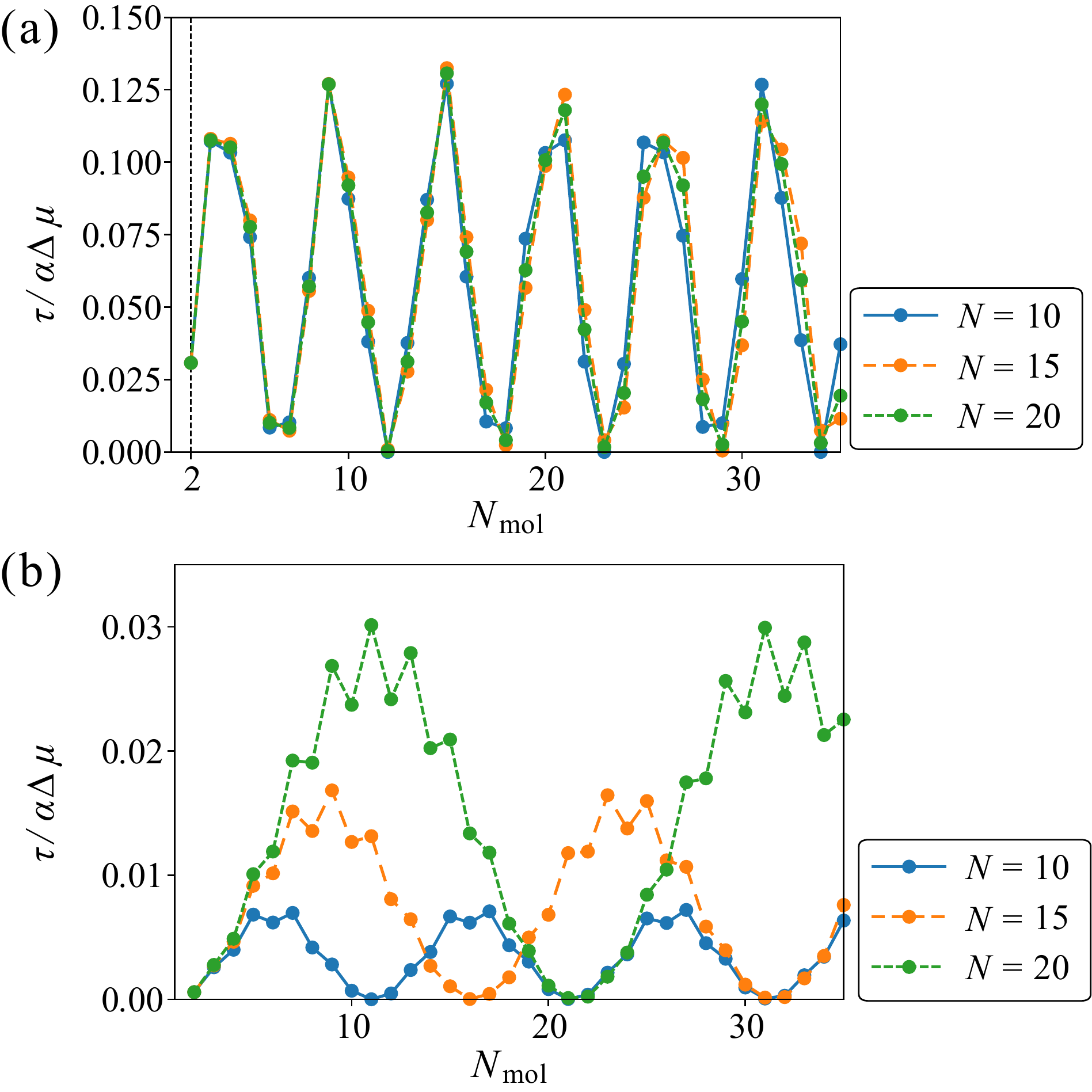}
\caption
{
The normalized torque as a function of the length of the molecule, for different lengths of the unit cell, $N=10, 15, 20$. (a) Strong SOI coupling,  $\alpha = 0.6$; (b) weak SOI coupling, $\alpha = 0.01$. For small   $\alpha$,  the period of the oscillations approaches  the length of the unit cell, see panel (b).
}
\label{fig:t_vs_n2}
\end{figure}

To further elaborate on this point, we display in Figs.~\ref{fig:t_vs_n2}  the normalized torque as a function   of the length of the molecule for several choices of the length of the unit cell. In panel (a) the SOI coupling is large, $\alpha=0.6$. The three curves (for three choices of the length of the unit cell)
 overlap each other.
Figure~\ref{fig:t_vs_n2} (b) shows the normalized torque for a  small SOI coupling strength, $\alpha=0.01$.
In this regime we find that the amplitude and the period do depend on the number of sites in the unit cell.
Furthermore, the period approaches  the length of the unit cell, i.e., the number of sites there,  when $\alpha$ tends to zero.

\begin{figure}[ht]
\includegraphics[width=.8 \columnwidth]{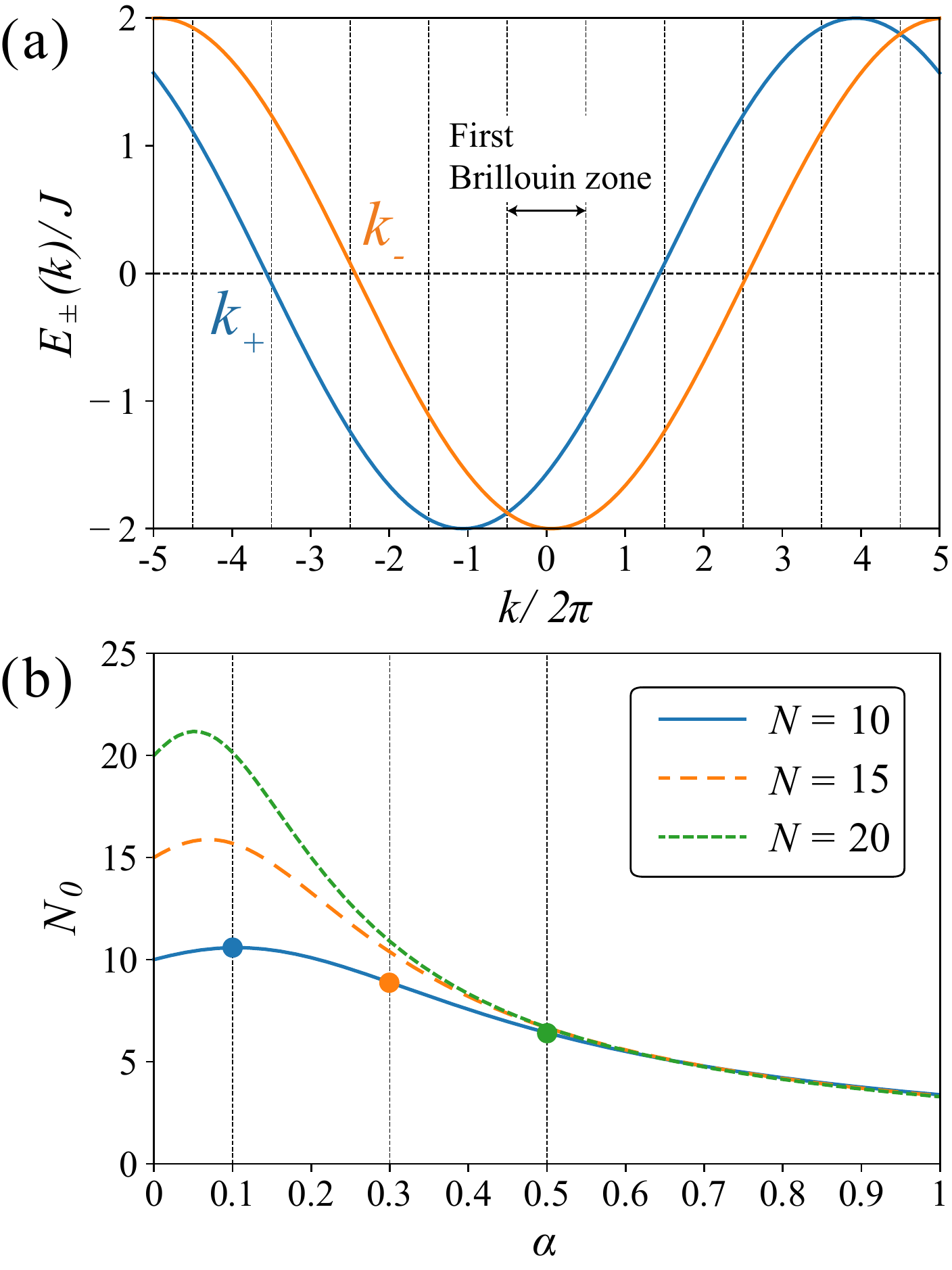}
\caption
{(a) The band structure of the DNA Hamiltonian Eq. (\ref{eq:H_mol}), in the extended Brillouin zone scheme for $\alpha=0.3$ [see Eq.~(\ref{eq:ene_dis1})].
The shift of two bands is caused by the SOI and the helical structure of the molecule.
(b) The period of oscillations in Fig.~\ref{fig:t_vs_n} as a function of the strength of the spin-orbit interaction for several values of the length of the unit cell $N$.
The dots represent values taken from Fig.~\ref{fig:t_vs_n}, for $\alpha=0.1,0.3, $ and 0.5. }
\label{fig:period_vs_soi}
\end{figure}

The origin of the oscillations of the mechanical torque,  depicted in Figs. \ref{fig:t_vs_e}, \ref{fig:t_vs_n},  and \ref{fig:t_vs_n2},  is the difference between the  wave numbers of the two spin states that correspond to the same energy. That is,
due to the SOI and the helical structure,  the  two spin states of two different wave numbers $k_+$ and $k_-$,   are degenerate. To further explore this point, we present in Appendix
\ref{sec:energy_band} a detailed calculation of the band structure resulting from our model Hamiltonian for the DNA molecule [see Eq. (\ref{eq:H_mol}) and the discussion leading to it]. Based on the expression derived in Appendix
\ref{sec:energy_band} for the energy dispersion, Eq.~(\ref{eq:ene_dis1}), we obtain the band structure displayed in  the extended zone scheme in
Fig.~\ref{fig:period_vs_soi} (a) with the onsite energy $\epsilon_{n}$ set to zero.
The first Brillouin zone is in the region $|k| <  \pi/N$.
The minima of  the $+$ and $-$ bands are shifted with respect to one another, and are located at $-\Delta k -2\pi/N$ and $\Delta k$, respectively.~\cite{Aronov1993}  [See Eq.~(\ref{eq:ene_dis}) for the explicit expression of $\Delta k$.]
As is shown in Appendix
\ref{sec:energy_band}, the period of the oscillations is
\begin{align}
N_0 = \frac{2 \pi}{k^{}_+ - k^{}_-} = \frac{\pi}{\Delta k}
\, .
\label{period}
\end{align}
where $\Delta k$ is given in Eq.
(\ref{eq:ene_dis}).
The period of the oscillations,   $N_0$,  as a function of the strength of the SOI $\alpha$ is plotted in
Fig.~\ref{fig:period_vs_soi} (b),  for several values of the length $N$ of the unit cell.
Within a reasonable parameter range, the period decreases as the  SOI strength increases, as  observed in Fig~\ref{fig:t_vs_n}.
For large SOI's values, the period becomes independent of the size of the unit cell.
It is worthwhile to note   that because of the helical structure of the molecule, a finite difference $k_+ - k_- = 2 \pi/N$ persists even in the limit $\alpha \to 0$  and thus even then the period approaches the length of the unit cell, i.e.,   $N_0 \to N$.
This explains the observations discussed in connection with Fig.~\ref{fig:t_vs_n2} (b).

\begin{figure}[ht]
\includegraphics[width=.8 \columnwidth]{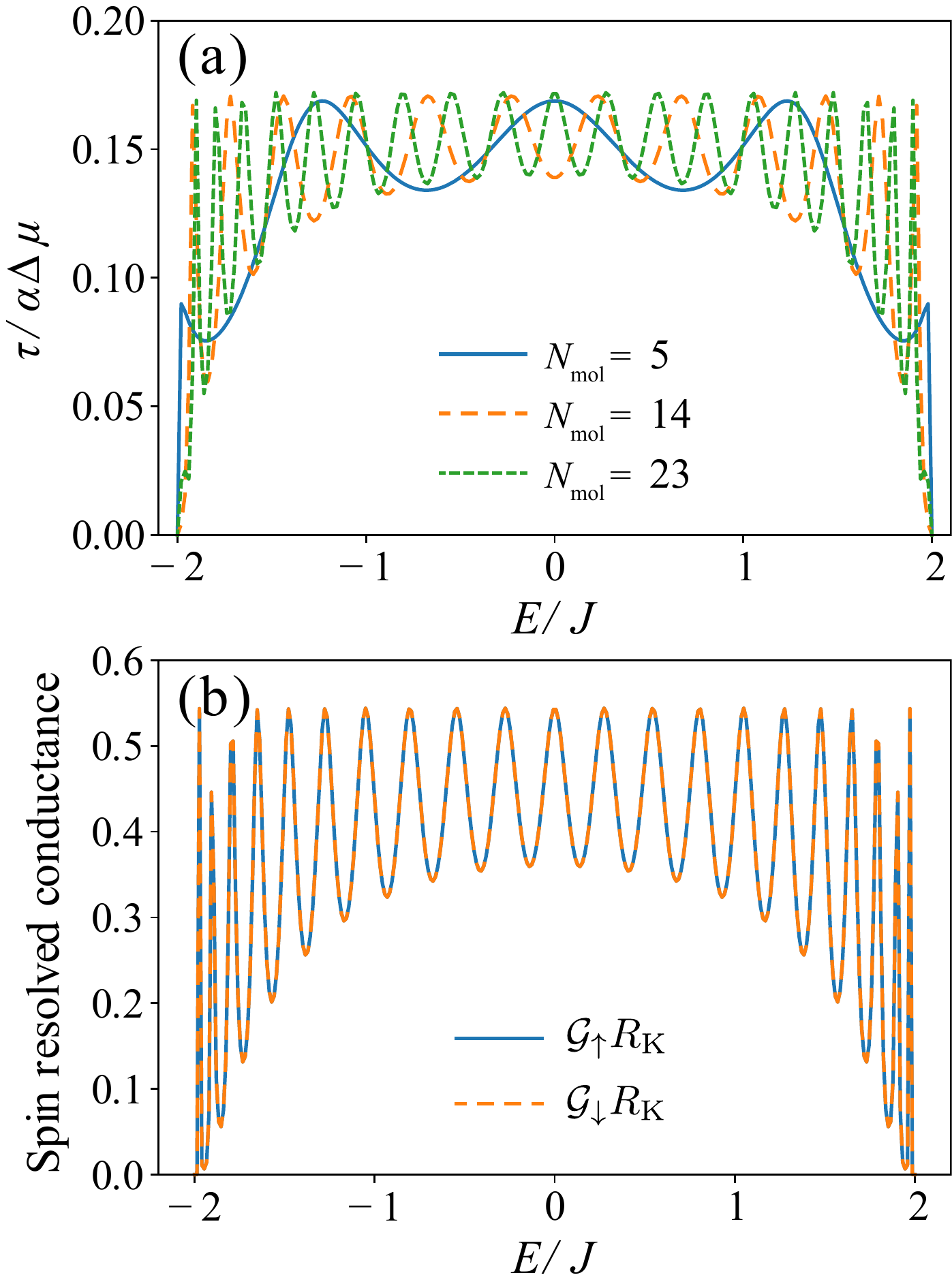}
\caption{
(a) The normalized mechanical torque as a function of the energy for  a molecule composed of  $N_{\rm mol}=5,14,23$ sites, for which  maximal values are reached in Fig.~\ref{fig:t_vs_n}.
The spin-orbit interaction strength is $\alpha=0.3$.
(b) The spin-resolved conductance as a function of the energy for $N_{\rm mol}=23$. The solid line is the spin-resolved conductance of an electron with spin $\uparrow$ and the dashed line is for an electron with spin $\downarrow$.
}
\label{fig:t_vs_ep}
\end{figure}

In Fig.~\ref{fig:t_vs_ep} (a) we draw the energy dependence of the mechanical torque for several lengths of the molecule. These correspond to the locations of the peaks   of the oscillations in Fig.~\ref{fig:t_vs_n} for $\alpha=0.3$.
Although the mechanical torque slightly oscillates as a function of the energy of the injected electrons, the average value can be as large as $\tau \sim 0.15 \alpha \Delta \mu$.
The magnitude of the mechanical torque and the resulting force acting on the DNA molecule may be estimated as follows.
The bandwidth of the DNA is approximately $4J \sim 120 \,{\rm meV}$, see Ref.~\onlinecite{Gutierrez2012}.
It follows that the maximal bias-voltage that can be applied  between the source and drain electrodes is $\Delta \mu \sim 4J$ and thus the maximum mechanical torque is roughly estimated as $\tau \sim 0.15 \alpha \Delta \mu \sim 8.7 \times 10^{-21} \,{\rm N \cdot m}$.
Such a value of the mechanical torque implies that the force acting on the molecule is
$F \sim \tau/R \sim  0.87 \, {\rm pN}$.
This value is not negligible  compared with the entropic elasticity force of a single double-stranded DNA, which is about $\sim 10 \, {\rm pN}$.~\cite{Bustamante2003, Bustamante1994}

It is a somewhat a delicate issue to estimate the value of the SOI coupling, $\alpha$. For a
B-form DNA, \cite{Bhushan2014} typical parameters are
$N \sim 10$, $R \sim 1$nm and $\Delta h \sim 3.4$nm.
The electric field acting on the electrons moving along the helical chain is approximately $E_0 \sim 4.5\times 10^{11}$V/m. \cite{Naaman2012}
However,  these values yield a rather  tiny SOI coupling,  $\alpha \sim 1.6 \times 10^{-4}$. Our calculations are performed for the range $0.01 \sim 0.6$ of $\alpha$ values,
which are about four orders of magnitude larger.
It is customary \cite{Guo2014,Matityahu2016} to adopt such values for the analysis of the effect of the chiral-induced spin selectivity. \cite{equation}

Figure~\ref{fig:t_vs_ep} (b) exhibits the energy dependence of the spin-resolved conductance for $N_{\rm mol}=23$ when an unpolarized electron is injected.
These spin-resolved conductances of the $\uparrow$-spin and $\downarrow$-spin are
\begin{subequations}
\begin{align}
\mathcal{G}^{}_\uparrow
R^{}_{\rm K}
&=
4\pi ( g^{}_{L\uparrow, R\uparrow} + g^{}_{L\uparrow, R\downarrow} )
\, ,
\\
\mathcal{G}^{}_\downarrow
R^{}_{\rm K}
&=
4\pi ( g^{}_{L\downarrow, R\uparrow} + g^{}_{L\downarrow, R\downarrow} )
\, ,
\end{align}
\end{subequations}
where $R_{\rm K} = h/e^2$ is the von Klitzing constant.
Since the spin-resolved conductance of the  $\uparrow$-spin is compatible with that of the $\downarrow$-spin, the output current is  not spin polarized.
This result can be deduced from the Bardarson theorem. \cite{Bardarson2008}
The theorem implies that for a single-channel two-terminal conductor which is time-reversal symmetric,  the spin-resolved transmission probability can be diagonalized in spin space by properly choosing the spin quantization axis.
Moreover, the up and down spin transmission probabilities are identical.
Therefore, if the injected electrons are (spin) unpolarized, it is not possible to induce a mechanical torque.

\section{Summary}
\label{sec:summary}

In the present paper, we have discussed the mechanical torque exerting on a spin-orbit interaction active helical DNA molecule induced by spin polarized currents that flow in response to a bias voltage.
The equation of motion [Eq. (\ref{eq:spin_conti_eq})] of a rotating molecule, which relates   the spin current and the mechanical torque $\tau$ in the steady state is derived, based on a specific microscopic Hamiltonian for the chiral molecule. The spin currents are expressed in a
Landauer-type formula, which involves
the spin-mixing spin-conductance, Eq.   (\ref{eq:sm-sc}). The latter  is an important ingredient of the  mechanical torque, which is generated as the back action of the spin-flip transmission and reflection processes.
It is found that the torque oscillates as a function of the length of the DNA molecule, with a period that is dominated by the strength of the spin-orbit interaction and the helix structure. 
When this interaction is very weak, the period approaches  the length of the unit cell

This work was supported by JSPS KAKENHI Grants 17K05575 and JP26220711,
by the Israeli Science Foundation (ISF), by
the infrastructure program of Israel Ministry of Science and Technology under contract 3-11173, and by a grant from
the Pazy foundation.

\begin{appendix}

\section{Expectation values of the  spin currents}
\label{sec:detailed_Landauer}

The expectation value of the spin-current operator,  Eq.~(\ref{eq:sco}),  in the scattering state excited by an electron impinging from the right [see  Eq.~(\ref{eq:ss})]  is
\begin{align}
&\langle I^{}_{r} \rangle^{}_{R,\ell \sigma}
\equiv
{}_R\langle \ell \sigma^{(+)}(\phi)| I^{}_{R} | \ell \sigma^{(+)}(\phi) \rangle^{}_R
\nonumber\\
=&
\sum_{\ell',\sigma'}
{}_R^{}\langle \ell \sigma^{(+)}(\phi)| \ell' \sigma' \rangle_r
\langle \ell' \sigma'  | I_{r} | \ell \sigma^{(+)} (\phi) \rangle_R \ ,
\label{0}
\end{align}
where $r=L,R$. In the second step we have neglected the states in the scattering region, i.e., the localized  states in the molecule (which are localized due to the  finite length of the molecule).
Exploiting the relation
\begin{align}
{}_r^{}\langle \ell' \sigma'  | I^{}_{r}
=
-\frac{i}{2} {}^{}_r\langle \ell' \sigma'  | V \sigma'
\ ,
\end{align}
where $\sigma'$ at the far end of the right hand-side of this equation is to be read as
$\sigma'=1 (-1)$ for $\sigma'=\uparrow (\downarrow)$,
we transform Eq.~(\ref{0}) into
\begin{align}
\langle I^{}_{r} \rangle_{R,\ell \sigma}
=&
-\frac{i}{2}
\sum_{\ell',\sigma'} \sigma'
{}^{}_R\langle \ell \sigma^{(+)} (\phi)| \ell' \sigma' \rangle^{}_r
\nonumber \\
&\times {}^{}_r\langle \ell' \sigma'  | V | \ell \sigma^{(+)} (\phi) \rangle^{}_R
\nonumber\\
=&
-\frac{i}{2}
\sum_{\ell',\sigma'} \sigma'
{}^{}_R\langle \ell \sigma^{(+)} (\phi)| \ell' \sigma' \rangle^{}_r
\nonumber \\
&\times {}_r\langle \ell' \sigma'  | T(E_{R,\ell \sigma}+i 0) | \ell \sigma \rangle^{}_R
\ .
\label{eq:1}
\end{align}
In the last step of Eq. (\ref{eq:1}) we have written the scattering state Eq.~(\ref{eq:ss})  using the $T$-matrix operator [see Eq.~(\ref{eq:t_matrix})],
\begin{align}
| \ell \sigma^{(+)} (\phi) \rangle^{}_R
=
[1 + G_0(E^{}_{R,\ell \sigma} + i0) T(E^{}_{R,\ell \sigma} + i0) ] | \ell \sigma \rangle^{}_R
\, ,
\label{eq:1a}
\end{align}
where the Green's function operator of the unperturbed system is
\begin{align}
G^{}_0(E) =  [E-\mathcal{H}_{\rm mol} - \mathcal{H}_{L} - \mathcal{H}_{R}]^{-1}_{}\ .
\end{align}
From Eq.~(\ref{eq:1a}) we obtain
\begin{align}
&{}^{}_R\langle \ell \sigma^{(+)} (\phi)| \ell' \sigma' \rangle^{}_r
=
\delta_{R,r} \delta_{\ell,\ell'} \delta_{\sigma,\sigma'}
\nonumber \\
&+
\frac{
{}_R\langle \ell \sigma| T(E_{R,\ell \sigma}-i0) | \ell' \sigma' \rangle_r
}{E_{R,\ell \sigma}-i0-E_{r,\ell' \sigma'}}
\, .
\label{eq:2}
\end{align}
Inserting Eq.~(\ref{eq:2}) into Eq.~(\ref{eq:1}) and  neglecting the imaginary part, the expectation value is
\begin{align}
\langle I^{}_{r} \rangle^{}_{R,\ell \sigma}
=&
\sigma
\frac{\delta_{R,r}}{2}
{\rm Im} {}_R\langle \ell \sigma  | T(E^{}_{R,\ell \sigma}+i 0) | \ell \sigma \rangle^{}_R
\nonumber \\
&+
\frac{\pi}{2}
\sum_{\ell',\sigma'} \sigma'
\delta(E^{}_{R,\ell \sigma}-E^{}_{r,\ell' \sigma'})
\nonumber \\ & \times
| ^{}_r\langle \ell' \sigma'  | T(E^{}_{R,\ell \sigma}+i 0) | \ell \sigma \rangle^{}_R |^2
\, .
\label{eq:3}
\end{align}
At this stage we utilize the optical theorem, \cite{JJSakurai}
\begin{widetext}
\begin{align}
{\rm Im} ^{}_R\langle  \ell \sigma  | T(E^{}_{R,\ell \sigma}+i 0) | \ell \sigma \rangle_R
=
- \pi
\sum_{r',\ell',\sigma'}
| ^{}_{r'}\langle \ell' \sigma'  | T(E^{}_{R,\ell \sigma}+i 0) | \ell \sigma \rangle^{}_R |^2
\delta(E^{}_{R,\ell \sigma}-E^{}_{r',\ell' \sigma'})
\, ,
\label{eq:ot}
\end{align}
where again the states in the scattering region are omitted.
Inserting Eq.~(\ref{eq:ot}) into the first term of Eq.~(\ref{eq:3}) yields
\begin{align}
\langle I^{}_{r} \rangle_{R,\ell \sigma}
=&
\sum_{r',\sigma'}
\frac{
\sigma' \delta^{}_{r',r}
-
\sigma
\delta^{}_{R,r}
}{2}
\pi
\sum_{\ell'}
| ^{}_{r'}\langle \ell' \sigma'  | T(E^{}_{R,\ell \sigma}+i 0) | \ell \sigma \rangle_R |^2
 \delta(E^{}_{R,\ell \sigma}-E^{}_{r',\ell' \sigma'})
\ .
\end{align}
For an electron whose energy is in the range $E < E_{R,\ell \sigma}<E+\delta E$,
where $\delta E$ is the level spacing in the right lead, the mechanical torque is
\begin{align}
\tau =
\sum_{r,r',\sigma'}
\frac{
\sigma' \delta^{}_{r',r}
-
\sigma
\delta^{}_{R,r}
}{2}
g^{}_{r \sigma', R \sigma}  \, \delta E
\ ,
\label{eq:4}
\end{align}
where $g^{}_{r \sigma', R \sigma}$ is the  spin-resolved spin-conductance,
\begin{align}
g_{r' \sigma', r \sigma}
=&
\pi
\sum_{\ell',\ell}
| ^{}_{r'}\langle \ell' \sigma'  | T(E_{r,\ell \sigma}+i 0) | \ell \sigma \rangle_{r} |^2
 \delta(E_{r,\ell \sigma}-E_{r',\ell' \sigma'})
 \delta(E_{r,\ell \sigma}-E)\nonumber\\
=&
\pi
\rho^{}_{r' \sigma'} (E)
| ^{}_{r'}\langle 1 \sigma'  | T(E+i 0) | 1 \sigma \rangle^{}_r |^2
\rho^{}_{r \sigma} (E)
\ ,
\end{align}
and the local density of states, $\rho^{}_{r \sigma} (E)$, is given in Eq. (\ref{eq:LDOS}).
\end{widetext}

Once the junction is biased, the electrons that contribute to the spin current are those whose energy $E$ is
 in the  window $\mu_L<E<\mu_L+\Delta \mu$, where $\mu_L$ is the chemical potential of the left lead.
Therefore, one can replace $\delta E$ in Eq.~(\ref{eq:4}) by the chemical potential difference $\Delta \mu$ and obtain Eq.~(\ref{eq:st}) in the linear-response regime and  at zero temperature.

In summary, the derivation in this Appendix is accomplished within the scattering formalism which pertains to noninteracting electrons. Other derivations, e.g.,  those based on the  many-particle wave function~\cite{Gurvitz1996} would result in the same expression. Likewise,
systematic approaches such as the Keldysh nonequilibrium Green's function technique~\cite{Schwabe1996,Utsumi1999} would produce the same results.

\section{The oscillations' periodicity}
\label{sec:energy_band}

As mentioned in the main text, the oscillations of the mechanical torque are related to the degeneracy of the spin states. Here we diagonalize the Hamiltonian of the molecule, Eq.  (\ref{eq:H_mol}),  to elaborate on this point. We use  periodic boundary conditions,  $c_j=c_{j+N_{\rm mol} }$. In this case the number of sites on the molecule,  $N_{\rm mol}$,  is a multiple of the number of sites in the unit cell, $N$, i.e.,  $N_{\rm mol} =MN$, where $M$ is a positive integer. Exploiting this observation, the
Hamiltonian (\ref{eq:H_mol})  takes the form~\cite{Matityahu2016} (the on-site energies are chosen to be zero)
\begin{align}
 \mathcal{H}^{}_{\rm mol} = -J \sum_{m=1}^M \sum_{n=1}^{N} c_{m,n+1}^\dagger V^{}_n c^{}_{m,n}  + {\rm H.c.}\ ,
\label{eq:H_mol_}
\end{align}
where  the  operators $c^{}_{n+Nm }$ ($c^{\dagger}_{n+Nm }$)  are written as $c^{}_{m,n}$ ($c^{\dagger}_{m,n}$). This form  satisfies the conditions
\begin{align}
c^{}_{m,n+N}&=c^{}_{m+1,n}\ ,\ \ \
c^{}_{m+M,n}=c^{}_{m,n}\ .
\end{align}
Introducing the discrete Fourier transform
\begin{align}
c_{\ell,n}^\dagger = \sum_{m=1}^M e^{i 2 \pi \ell m/M} c_{m,n}^\dagger/\sqrt{M}
\ ,
\end{align}
the Hamiltonian
(\ref{eq:H_mol_}) becomes
\begin{align}
{\mathcal H}^{}_{\rm mol} = \sum_{\ell=-M/2}^{M/2-1} {\mathcal H}^{}_{\rm mol}(\ell) \, ,
\end{align}
where the Bloch Hamiltonian ${\mathcal H}_{\rm mol}(\ell)$ is
\begin{align}
{\mathcal H}_{\rm mol}(\ell) =& -J \sum_{n=1}^{N-1} c_{\ell,n+1}^\dagger V^{}_n c^{}_{\ell,n} \nonumber\\
&- J c_{\ell,1}^\dagger V^{}_N c^{}_{\ell,N} e^{-i 2 \pi \ell/M}
+ {\rm H.c.} \, .
\label{eq:H_mol_1}
\end{align}
The matrix part of the tunneling amplitude, Eq.
(\ref{eqn:V_n}) (which is a unitary matrix) can be written in the form
\begin{align}
V^{}_n
=&\exp[
-i ((n+1/2)\Delta \varphi + \pi/2 + \phi ) \sigma_z /2]
e^{ i \alpha \bm{\hat{n}} \cdot \bm{\sigma} }
\nonumber \\
& \times
\exp[i ((n+1/2)\Delta \varphi + \pi/2 + \phi ) \sigma_z /2]
\, ,
\end{align}
where the unit vector $\hat{\bm n}$ is defined by this relation.
Choosing for simplicity the case where
the rotation angle is $\phi=-\Delta \varphi/2-\pi/2=-\pi/N-\pi/2$, the tunneling matrix becomes
\begin{align}
V^{}_n
=
e^{-i n \Delta \varphi \sigma_z /2}
e^{ i \alpha \bm{\hat{n}} \cdot \bm{\sigma} }
e^{i n \Delta \varphi \sigma_z /2}
\ ,
\label{eq:V_n_}
\end{align}
with the unit vector $\hat{\bm n}$  given by
\begin{align}
\bm{\hat{n}} &= [\sin (\theta), 0, \cos (\theta)]_{}^T\ ,
\nonumber\\
\tan (\theta )&= - \frac{\Delta h \Delta \varphi}{4 \pi R \sin(\Delta \varphi/2)} \ .
\end{align}
It is now expedient to rotate the spin quantization-axis at each site on the molecule by transforming the operators there,
\begin{align}
{c}^{}_{\ell,n}=e^{i 2 \pi \ell n/(MN)} e^{-i n \Delta \varphi (\sigma_z- {\bm 1}) /2}  \widetilde{c}^{}_{\ell,n} \, .
\end{align}
(note that the rotated operator obeys the periodic boundary conditions,
$\widetilde{c}^{}_{\ell,n+N}=\widetilde{c}^{}_{\ell,n}$.)
This rotation enables us to diagonalize the Hamiltonian
(\ref{eq:H_mol_1})  by exploiting  the discrete Fourier transform
\begin{align}
{\widetilde{c}_{\ell,p}}^\dagger = \sum_{n=1}^{N} e^{i 2 \pi p n/N} {\widetilde{c}_{\ell,n}}^\dagger/\sqrt{N}\ ,
\end{align}
which yields
\begin{align}
{\mathcal H}^{}_{\rm mol}(\ell) = \sum_{p=0}^{N-1} {\widetilde{c}_{\ell,p}}^\dagger {\cal H}^{}_p(\ell) {\widetilde{c}^{}_{\ell,p}} \, ,
\end{align}
where
\begin{align}
{\cal H}^{}_p(\ell) 
=&-J\exp\Big [-i\Big (\frac{(2p+1)\pi}{N}+k(\ell)\Big )\Big ]\nonumber\\
&\times\exp\Big [i \pi\sigma_z/N\Big ] \exp\Big [i \alpha  \bm{\hat{n}} \cdot \bm{\sigma}\Big ]+{\rm H.c.}\ .
\end{align}
Here we have denoted
$k(\ell)=2 \pi \ell/M$.  Introducing  the notations
\begin{align}
\exp\Big [i \pi\sigma_z/N\Big ] \exp\Big [i \alpha  \bm{\hat{n}} \cdot \bm{\sigma}\Big ]
=B_0 {\bm 1} + i \, \bm{\sigma} \cdot \bm{B}\ ,
\end{align}
where
$B_0^2+|{\bm B}|^2=1$,
we obtain
\begin{align}
{\cal H}^{}_p(\ell) =&
-2 J \Big [ B_0 \cos \Big (\frac{(2p+1)\pi}{N}+k(\ell)\Big )
\nonumber \\
& - {\bm \sigma} \cdot {\bm B} \sin \Big (\frac{(2p+1)\pi}{N}+k(\ell)\Big )\Big ]
\, ,
\label{Hpl}
\end{align}
where  
\begin{align}
B^{}_0
=&
\cos \left(\frac{\pi}{N} \right) \cos (\alpha  )-
\cos( \theta )\sin \left(\frac{\pi}{N} \right) \sin (\alpha) \, , \nonumber\\
B^{}_x
=&
\sin (\theta) \cos \left(\frac{\pi}{N} \right) \sin (\alpha) \, ,
\nonumber\\
B^{}_y
=&
-\sin (\theta )\sin \left(\frac{\pi}{N} \right) \sin (\alpha) \, ,
\nonumber\\
B^{}_z
=&
\cos (\theta )\cos \left(\frac{\pi}{N} \right) \sin( \alpha)
+
\sin \left(\frac{\pi}{N} \right) \cos (\alpha)
\, .
\label{B}
\end{align}
It is rather straightforward \cite{Aronov1993} to diagonalize the Bloch Hamiltonian ${\cal H}^{}_{p}(\ell)$, Eq. (\ref{Hpl}).
The eigenvalues are
\begin{align}
E^{}_{p,\pm}[k(\ell)]=-2J\cos
\Big (\frac{(2p+1)\pi}{N}+k(\ell)\pm\Delta k\Big ) \ ,
\label{eq:ene_dis0}
\end{align}
where the phase shift $\Delta k$  is
\begin{align}
\Delta k = \arctan \left( \frac{|{\bm B}|}{B_0} \right) \, .
\label{eq:ene_dis}
\end{align}
It is interesting to note that in the absence of the SOI, i.e., when $\alpha=0$, the phase shift is $\Delta k=\pi/N$ [see Eqs.
(\ref{B})] and the eigen energies become $
E^{}_{p,+}[k(\ell)]=
 -2J \cos [k(\ell)/N +2 \pi (p+1)/N]$ and
 $E_{p,-}[k(\ell)] = -2J \cos [k(\ell)/N +2 \pi p/N]$.
That is, even without the SOI, the energy dispersions of up and down spins do not coincide for a given band index $p$, $E_{p,+}[k(\ell)] \neq E_{p,-}[k(\ell)]$.

 In general, there are four wave numbers for a given energy $E$,  associated with left- and right-moving electrons and  the two spin components.
 The wave vectors corresponding to the left-moving electron are
 \begin{align}
k^{}_\pm = \arccos \left( - \frac{E}{2J} \right) \mp \Delta k - \frac{\pi}{N}
 \ ,
 \end{align}
 where in the extended-zone scheme
\begin{align}
E_\pm(k) = -2J \cos (k/N + \pi/N  \pm \Delta k)
\, .
\label{eq:ene_dis1}
\end{align}
Consequently, the period of $k$ is $2\pi N$.
 Since the two spin wave functions  of the propagating electron have different wave vectors, the propagation is accompanied by interference of the two wave functions, i.e., by spin precession. The periodicity of the resulting oscillations, $N_{0}$ (in our case, in the mechanical torque) is determined by the difference between the two wave vectors, that is, 
\begin{align}
N_0 = \frac{2 \pi}{k^{}_+ - k^{}_-} = \frac{\pi}{\Delta k}
\tag{\ref{period}}
\, .
\label{eq:period}
\end{align}

\end{appendix}


\begin{thebibliography}{99}
\bibitem{AAcom} Interestingly, $[\hat{\bm K}^{}_{n+1,n} \times {\bm \sigma}]^{}_z=-(\lambda/\alpha) (\Delta h/N){\bm E}^{}_{n+1,n}\cdot {\bm \sigma}$, which shows that the torque vanishes when the pitch vanishes. The helicity is crucial.
\bibitem{equation}
In Ref.~\onlinecite{Matityahu2016}, the strength of the SOI is characterized by a parameter $\theta$.
It is related to our parameters by
\begin{align}
\theta &= N \arccos [ \cos (\pi/N) \cos (\alpha )\nonumber\\
&+ 2(R/\ell) \sin^2 (\pi/N) \sin( \alpha )] +(2\tilde{n}+1)\pi
\nonumber
\ ,
\end{align}
 where $\tilde{n}$ is an integer.
In our case, for $\Delta h/R=18.1$, $N=10$ and $\alpha=0.3$,
we find $\theta/\pi = 0.12 + 2\tilde{n}$, which is about  the same order of magnitude as the values quoted  in Ref.~\onlinecite{Matityahu2016}.
\end{thebibliography}
\end{document}